\documentclass[showpacs,twocolumn,aps]{revtex4}
\usepackage{epsfig}

\newcommand{\nc}{\newcommand}
\nc{\be}{\begin{equation}}
\nc{\ee}{\end{equation}}
\nc{\bea}{\begin{eqnarray}}
\nc{\eea}{\end{eqnarray}}
\nc{\disp}{\displaystyle}
\nc{\ade}{\mbox{$A$-$D$-$E$}}
\nc{\calN}{{\cal N}}
\nc{\calC}{{\cal C}}
\nc{\calM}{{\cal M}}
\nc{\calS}{{\cal S}}
\nc{\phit}{\hat{\varphi}}
\nc{\chit}{\hat{\chi}}
\nc{\hcalN}{\hat{\calN}}
\nc{\hcalS}{\hat{\calS}}
\nc{\hS}{\hat{S}}
\nc{\sigmad}{\sigma^\dagger}
\nc{\psid}{\psi^\dagger}

\font\tenmsb=msbm10
\font\sevenmsb=msbm7
\font\fivemsb=msbm5
\newfam\msbfam
\textfont\msbfam=\tenmsb
\scriptfont\msbfam=\sevenmsb
\scriptscriptfont\msbfam=\fivemsb

\def\bra#1{\langle #1|}
\def\ket#1{|#1\rangle}
\def\e{{\rm e}}
\def\d{{\rm d}}
\def\i{{\rm i}}

\begin{document}
\title{Stochastic processes and conformal invariance}

\author{Jan de Gier$^{1}$, Bernard Nienhuis$^2$, Paul A. Pearce$^1$ and
Vladimir Rittenberg$^{1,3}$
}
\affiliation{
{}$^1$Department of Mathematics and Statistics,
University of Melbourne, Parkville, Victoria 3010, Australia\\
{}$^2$Universiteit van Amsterdam, Valckenierstraat 65, 1018 XE
Amsterdam, The Netherlands\\
{}$^3$Physikalisches Institut, Bonn University, 53115 Bonn, Germany.
}

\date{\today}

\begin{abstract}
We discuss a one-dimensional model of a fluctuating interface with 
a dynamic exponent $z=1$. The events that occur are adsorption, which 
is local, and desorption which is nonlocal and may take place over regions
of the order of the system size. In the  thermodynamic limit, the time 
dependence of the system is given by characters of the $c=0$ logarithmic conformal
field theory of percolation. This implies in a rigorous way a
connection between logarithmic conformal field theory 
and stochastic processes. The finite-size scaling behavior of the
average height, interface width and other observables are
obtained. The avalanches produced during desorption are analyzed and
we show that the probability distribution of the avalanche sizes obeys
finite-size scaling with new critical exponents. 
\end{abstract}

\maketitle
The structure of growing interfaces continues to be a subject of 
major interest and a characterization of the various universality 
classes of critical behavior remains an open question \cite{BaraS95}. 
In this paper we present a one-dimensional adsorption-desorption model of 
a fluctuating interface which belongs to a new universality class where the 
dynamical critical exponent $z=1$. In this model the interface evolves 
following nonlocal Markovian dynamics (for an adsorption-desorption model 
with local dynamics and $z>2$ see \cite{KhodD97}). The relaxation rules are such 
that one has avalanches with a long tail in their probability distribution 
function (PDF). The model belongs therefore to the self-organized criticality 
class (SOC) \cite{BakTW87,Jensen98,BenHur96}. What makes the model
special is that the correlation lengths in the time and space
directions are both proportional to the size  of the system (this is
not the case for other SOC models \cite{Jensen98,BuldHS93}). The
single scale that exists in the system is therefore its
size. Moreover, our model has the big advantage of being solvable. 
 
The Hamiltonian which gives the time evolution of our model is
integrable. The finite-size scaling (FSS) limit of its spectrum can be obtained
using the Bethe Ansatz \cite{AlcarazBBBQ} and is given by characters of a $c=0$ 
Virasoro algebra. A logarithmic conformal field theory (LCFT) with
$c=0$ \cite{KogN02} appears also in other domains of physics such as
systems with quenched disorder and the  quantum Hall effect
\cite{GurL99}, and possibly string theory \cite{KogP01}. 
Moreover \cite{GierPRN}, the PDF describing the stationary state is related 
to combinatorial aspects of the ice model defined on a rectangle with 
special boundary conditions. This is probably the reason why we are
able to conjecture exact expressions for various quantities
connected to the model even for finite lattices.
 
We consider an interface on a one-dimensional lattice of size 
$L+1$ ($L=2n$ even). The non-negative heights $h_i\; (i=0,1,..,L)$
that specify the interface, obey restricted solid-on-solid (RSOS) 
rules,
\be
h_{i+1}-h_i = \pm 1,\quad h_0=h_L=0,\quad h_i \geq 0.
\ee
Alternatively, we can describe the interface using slope variables 
$s_i= (h_{i+1}-h_{i-1})/2$, ($s_0=s_L=0$). The interface evolves
through adsorption and desorption according to the following
rules. Adsorption, which locally changes the height $h_i$ to $h_i+2$,
takes place with a rate equal to $1$ at a local  
minimum of the interface ($s_i=0$, $h_i < (h_{i-1},h_{i+1}$)). The non-vanishing 
rates for desorption can be understood using the notion of active segments. 
A segment of a configuration with the end-points $a$ and $b$ is 
defined by the conditions: $h_a=h_b=h$ and $h_i > h$ for $a<i<b$. 
We call a segment active if at least one of the boundary slopes $s_a$
or $s_b$ is nonzero. In the desorption event, all the $b-a-1$
heights $h_i$ contained in a  segment decrease by two units ($h_i
\mapsto h_i-2$) with a
rate
\be
\delta(s_a-1)+\delta(s_b+1), \label{eq:rates}
\ee
where $\delta$ is the discrete Kronecker symbol, the other heights remaining unchanged.
In order to find which desorption events can take place, for each of the
$C_n=(2n)!/((n+1)(n!)^2)$ configurations one first looks at how many
active segments one has and then one uses (\ref{eq:rates}).  
If the two slopes $s_a$ and $s_b$ are zero, the desorption rate is zero.
This observation has two consequences. First, in the stationary state 
we expect to see predominantly configurations with large terraces, i.e.
intervals where the slope is zero for all the sites. Next, it is  
meaningful to consider clusters. Those are segments where the end-points 
have the heights equal to zero. According to the rules (\ref{eq:rates}) 
desorption takes place only within a cluster.

The dynamics of the interface can easily be visualized using tiles
(tilted squares) which cover the area between the interface and the 
substrate ($h_{2i}=0$, $h_{2i+1}=1$). For a given configuration $c$, the 
number of tiles is $u(c)=\frac12\sum_{i=2}^{L-2} h_i+1-L/4$. According to 
the rules given above, through adsorption one adds one tile, through 
desorption one loses a layer with $b-a-1$ tiles (this is an odd number). 
If one looks at our model from the point of view of self-organized critical 
phenomena, the number of tiles $u$ lost in a desorption event 
defines the size of an avalanche.

The physics of our model can be understood in the following way. We
have a gas of tiles in the presence of a surface. Tiles uniformly try
to attach to the interface and are succesful on sites that are lower
than their neighbors. Tiles get reflected back into the gas on sites
that are higher than their neighbors and trigger desorption events and
are reflected on local slopes. Because the stationary state mainly
consists of terraces, desorption events occur less frequently, but
when they occur, they can take larger number of tiles. 
 
Before describing the physical properties of the model we will give 
another formulation which enables us to obtain the FSS spectrum of the 
Hamiltonian, defined by the rates given above and thus to find the 
dynamic exponent. To each RSOS configuration with $L+1$ sites one can 
associate a configuration of 
$n$ non-intersecting half-loops on $L=2n$ sites. The height $h_i$ is the 
number of half-loops above the midpoint of the sites $i$ and $i+1$ in 
the half-loop configuration. For example, the following picture illustrates 
the association between a configuration with three half-loops and an
$L=6$ RSOS configuration with two clusters and one tile,
\be
\begin{picture}(20,5)
\put(0,0){\epsfxsize=1in\epsfbox{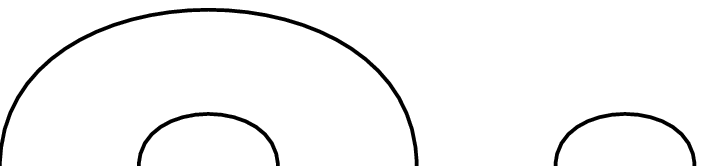}}
\put(-0.5,-3){$\scriptstyle 1$}
\put(3.4,-3){$\scriptstyle 2$}
\put(7.3,-3){$\scriptstyle 3$}
\put(11.2,-3){$\scriptstyle 4$}
\put(15.1,-3){$\scriptstyle 5$}
\put(19,-3){$\scriptstyle 6$}
\end{picture}
\quad = \quad
\begin{picture}(22,4)
\put(0,0){\epsfxsize=1in\epsfbox{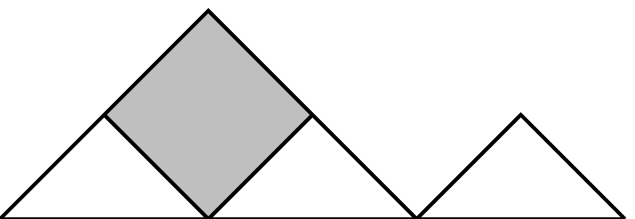}}
\put(-0.5,-3){$\scriptstyle 0$}
\put(2.8,-3){$\scriptstyle 1$}
\put(6.1,-3){$\scriptstyle 2$}
\put(9.4,-3){$\scriptstyle 3$}
\put(12.7,-3){$\scriptstyle 4$}
\put(16,-3){$\scriptstyle 5$}
\put(19.2,-3){$\scriptstyle 6$}
\end{picture}.
\label{dyck}  
\ee
\vskip2mm

The half-loop configurations on the other hand, can be associated with 
the left ideal of the Temperley Lieb (TL) algebra at $Q=1$
\cite{Mart90}. The $L-1$ generators of the algebra satisfy the relations
\be
e_i^2= e_i,\; e_ie_{i\pm 1}e_i=e_i,
\; [e_i,e_j]=0,\  |i-j|>1,
\ee
and have the graphical representation
\be
e_i \quad =\quad 
\begin{picture}(40,5)
\put(0,-2){\epsfxsize=2in\epsfbox{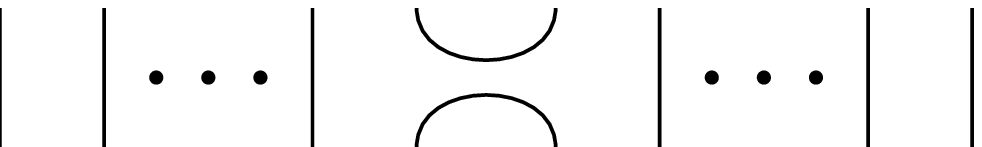}}
\put(-.5,-5){$\scriptstyle 1$}
\put(4,-5){$\scriptstyle 2$}
\put(11.5,-5){$\scriptstyle i-1$}
\put(17,-5){$\scriptstyle i$}
\put(20,-5){$\scriptstyle i+1$}
\put(25,-5){$\scriptstyle i+2$}
\put(33,-5){$\scriptstyle L-1$}
\put(39.5,-5){$\scriptstyle L$}
\end{picture}\;,
\label{monoid}
\ee
\vskip5mm
\noindent
The left ideal is generated by the action of TL generators on $I_0 = 
\prod_{i=1}^{n} e_{2i-1}$. For example the action of $e_3$ on the
half-loop configuration (\ref{dyck}) is
\be
\begin{picture}(50,4)
\put(0,-4){\epsfxsize=1in\epsfbox{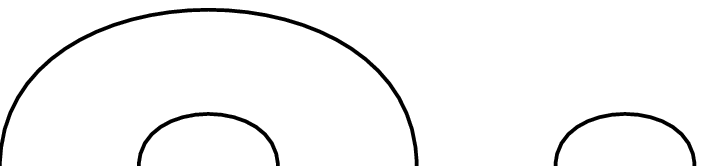}}
\put(23,0){$=$}
\put(28,0){\epsfxsize=1in\epsfbox{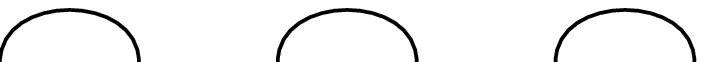}}
\end{picture}\;.
\ee
\vskip.1in
The Hamiltonian gives the time evolution in the vector space of the
half-loop (RSOS) configurations,
\be
H=\sum_{j=1}^{L-1} \left(1-e_j\right),\quad
\frac{\d}{\d t} P_c(t) = - \sum_d H_{cd} P_d(t).
\label{eq:ham}
\ee
$P_c(t)$ is the (unnormalized) probability to find the system in the 
configuration $c$ at the time $t$ and $r_{cd} = -H_{cd}$ give the rates 
of our model for the transitions $d \rightarrow c$. Since $H$ is an intensity matrix 
($\sum_c H_{cd} =0$) \cite{GierPRN}, it has a zero eigenvalue with a 
trivial bra and a nontrivial ket which gives the probabilities in the 
stationary state,
\be
\begin{array}{l}
\displaystyle \bra0\,H=0,\quad
\bra0=(1,1,\ldots,1), \\[3mm]
\displaystyle
H\ket0 = 0,\quad \ket0=\sum_c P_c\ket c,\quad P_c=\lim_{t\to\infty}
P_c(t).
\end{array}
\ee

Let us compute the dynamic critical exponent $z$. We use the following 
representation of the TL algebra
\bea
e_i&=& \frac14 - \frac12 \left[ \Big(\sigma^x_i\sigma^x_{i+1} +
\sigma^y_i\sigma^y_{i+1} - \Delta\sigma^z_i\sigma^z_{i+1} \Big)
\right.\nonumber\\
&&\left.\mbox{} + h\Big(\sigma^z_i-\sigma^z_{i+1}\Big)\right], 
\label{eq:XXZrep}
\eea
where $\sigma^x,\sigma^y,\sigma^z$ are Pauli matrices, 
$-2\Delta=q+q^{-1}=1$, $2h=q-q^{-1}=
\i \sqrt{3}$ and $q=\e^{\pi\i/3}$.
In this representation $H$ given by (\ref{eq:ham}) becomes the Hamiltonian 
of the XXZ quantum chain \cite{PasqS90} with $L$ sites and the energy gaps 
$E_k$ (the ground-state energy is zero for any $L$) scale like
\be
E_k = \pi v_{\rm s} x_k L^{-z},
\ee
with $z=1$, $v_{\rm s}= 3\sqrt{3}/2$ is the sound velocity \cite{AlcarazBBBQ} 
and $x_k$ are related to surface exponents of a $c=0$ LCFT of
percolation \cite{ReadS01}. In the continuum limit, the spectrum of the
Hamiltonian (\ref{eq:ham}), and thus the values of $x_n$, in the full 
TL algebra is given by the Virasoro characters \cite{ReadS01}
\be
\chi_s(w) = w^{s(2s-1)/3}(1-w^{2s+1})\prod_{k=1}^\infty (1-w^k)^{-1}.
\ee
Here $w=\exp(\pi T v_{\rm s}/L)$ parametrizes the temporal ($T$) and spatial ($L$)
extent of the stochastic process. In the subspace of the RSOS
configurations the spectrum is given by $\chi_0(w)$. To our knowledge,
for the first time, a connection is made between stochastic processes
and LCFT. This implies among other things that the space and time
correlation lengths are the same and that, in the continuum limit, the
forms of the space-time correlation functions in the stationary state
are known \cite{CFT}.

As discussed in detail in \cite{GierPRN} the stationary state of our model 
with $L$ sites is related to the two-dimensional ice model \cite{ice} 
defined on a rectangle of dimension $L\times (L-1)/2$ with special
boundary conditions \cite{Kore82,Kupe96}. This model is equivalent to
a fully packed loop (FPL) model \cite{BatchBNY96} on the rectangle,
all configurations being equally probable. 
We briefly explain this connection and its consequences. If we
choose the (unnormalized) probability of the ``pyramid" configuration
($s_1=...=s_{n-1}=1$, $s_n=0$, $L=2n$) to be equal to one (this
configuration has the smallest probability), then the normalization
factor of the stationary state \cite{BatcGN01} is equal to 
\be
\bra00\rangle = \prod_{j=0}^{n-1}\, (3j+2) {(2j+1)!(6j+3)! \over
(4j+2)!(4j+3)!},
\ee
This is precisely $A_{\rm V}(2n+1)$, the number of configurations of
the FPL model on the rectangle. Moreover, the (unnormalized)
probabilities for the other half-loop configurations are integer
numbers which are equal to the numbers of configurations of the FPL
model with the external occupied edges connected in the same way as in
the configuration of half-loops \cite{RazS01,GierPRN}. This implies
that the PDF describing the stationary distribution in the RSOS
configurations space corresponds to a uniform probability distribution
in the space of the FPL model, or equivalently the ice model. The RSOS
configuration with the largest (unnormalized) probability is the one
with the longest terrace ($s_1=..=s_L=0$) in which the interface
coincides with the substrate. The normalized PDF is obviously
$p_c=P_c/\bra00\rangle$.
 
We now describe some properties of the model. They were obtained either 
numerically for lattice sizes up to $L=18$ or an exact expression was
conjectured and checked up to $L=18$. We start with the properties of
the stationary state. The fraction of the interface covered by
terraces \cite{Strog} is
\be
\tau(L) = \frac{1}{L-1} \sum_{j=1}^{L-1} \langle 1-|s_j|\rangle =
\frac{3L^2-2L+2}{(L-1)(4L+2)}.
\ee
This implies that for large $L$, three quarters of the surface is
covered by terraces. We now consider the number of clusters per
configuration $C = \sum_{j=1}^L \delta(h_j)$. Its average and the
average of its square have the following expressions 
\bea
\langle C\rangle &=& \frac13\prod_{j=0}^{n-1}
\frac{(2j+1)(3j+4)}{(j+1)(6j+1)}- \frac13 \approx 0.738 L^{2/3},\quad\\
\langle C^2\rangle &\approx& (1.29 -0.49 L^{-0.63})\langle C\rangle^2. \label{eq:c2}
\eea
From (\ref{eq:c2}) which
we deduce that the compressibility \cite{Pathria72} of the gas of
clusters $\kappa\approx 0.29L$ diverges like the size of the system. 

The average height $\langle \overline{h} \rangle$ and interface width $w_\infty$, 
which characterizes the roughness of the surface in the stationary state,
\be
\overline{h^m} = \frac{1}{L} \sum_{i=1}^L h_i^m,\quad w_\infty = 
\sqrt{ \langle \overline{h^2} - \overline{h}^2 \rangle}, 
\ee
are compatible with the following behavior,
\be
\langle \overline{h}\rangle \approx 0.14 \ln \left(L/2\right),\quad
w_\infty \approx 0.34 \left(\ln L/2\right)^{1/3}. \label{eq:heightfit} 
\ee
This implies that the PDF of the heights has a very small dispersion and 
also that the usual exponents \cite{BaraS95} $\alpha=\beta$ vanish.
As in any fit involving logs, the formulas given in (\ref{eq:heightfit}) 
are probably not the last word. What is certain is that the width grows 
slowly with the size of the system implying that the surface is only 
marginally rough. It is interesting to mention that marginally rough 
surfaces (with $z=1.581$ corresponding to the directed percolation 
universality class) were also encountered \cite{AlonEHM98}
at a critical point dividing a moving rough KPZ phase from a smooth, 
massive phase. In these models a factor $3$ between the powers of the
logs (like in  (\ref{eq:heightfit})) was also seen \cite{Hinri01}. The
factor $3$ is typical of a large class of growth problems
\cite{Forr01}. To complete the interface growth picture, we calculated
the ratio of the time dependent interface width $w(t,L)$ to
$w_{\infty}$ for various lattice sizes. The curves were obtained by
solving on a computer the differential equations (\ref{eq:ham}) taking the
substrate as the initial configuration. If $z=1$ and if the
Family-Vicsek scaling applies, this ratio should converge to a
function of $t/L$. We have checked that this is indeed the case. 
 
We now consider the avalanches in our model. If in the stationary 
state one considers the configuration $d$ with $u(d)$ tiles, it
changes with rate $r_{cd}$ into the configuration $c$ with $u(c)$ tiles. 
The rate of changes in which $u$ tiles are lost, can be written as
\be
R(u,L) = \sum_{c\neq d}\delta(u(d) - u(c)-u) r_{cd} p_d.
\ee
We have found the following values for the rate of adsorption of one
tile, $R(u=-1,L)$, and the rate of all possible avalanches, $R(u>0,L)$,
\bea
R(-1,L) &=& \frac{3L(L-2)}{4(2L+1)},\\
R(u>0,L) &=& L-2-2R(-1,L). \label{eq:tileabs}
\eea
Notice that for large $L$ the rate for the adsorption of
one tile ($u=-1$) is $3/2$ times larger than the rate for an
avalanche ($u>0$). This explains the relative rarity of the desorption
events.

\begin{figure}
\begin{picture}(56,40)
\put(0,0){\epsfxsize=2.8in\epsfbox{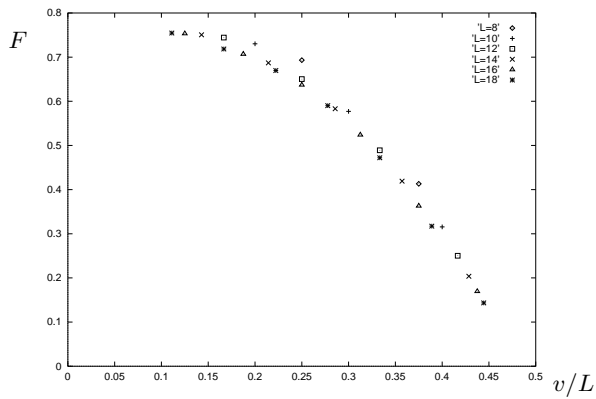}}
\put(-1,35){$F$}
\put(56,-1){$v/L$}
\end{picture}
\caption{Avalanche scaling function $F(v/L)$. The data are obtained for
$v>1$ and $L=8,\ldots,18$.}
\label{fig:avscale}
\end{figure}

Since through desorption one loses an odd number of tiles, it is 
convenient to write $u=2v-1$ and to consider $v$ as the size of an
avalanche. Given the occurrence of an avalanche, its size $v$ is
distributed according to the PDF
\be
S(v,L) = \frac{R(2v-1,L)}{R(u>0,L)}.
\ee
FSS \cite{Jensen98} predicts the following form for this PDF,
\be
S(v,L) = v^{-\tau} F(v/L^D). \label{eq:AvPDF}
\ee
One way to get the exponents of the FSS function is to consider the 
moments \cite{EvertszM},
\be
\langle v^k\rangle = \sum_{v=1} v^k S(v,L) \sim L^{\sigma(k)},
\ee
and one expects
\be
\sigma(k)=
\left\{ \begin{array}{cc}
0, & k<\tau-1 \\ 
D(k+1-\tau), & k > \tau-1
\end{array}\right. . 
\ee
From equation (\ref{eq:tileabs}) one gets $\langle v\rangle =
(5L+4)/4(L+2)$, which implies that for large $L$ the average size of
an avalanche is $3/2$ tiles. A numerical investigation of the other
moments for $1\leq k \leq 5$ indicates $\sigma(k)=k-2$. This implies 
$D=1$ and $\tau = 3$. 
The numerics can be quite precise even if we have data for $L$ up to
$18$. This is because we know $S(v,L)$ exactly and we can use VBS approximants
\cite{VBS} to derive the large $L$ behavior of the moments (the convergence is
less good around $\sigma(k)=0$ due to logarithmic effects). The value
$D=1$ was to be expected since $L$ is the only characteristic length in our 
system. A consistency check was done assuming $D=1$ in (\ref{eq:AvPDF}) to see for 
which value of tau one gets data collapse for the scaling function $F(v/L)$. Since we
have data up to $L=18$ only we cannot expect a precise value of $\tau$
nor of $F$. Nevertheless as shown in Fig. \ref{fig:avscale}, a data collapse is 
visible for $\tau=3.2$. 

To conclude, we have presented an SOC model of a critical fluctuating 
interface belonging to a new universality class. We have also 
shown the connection between the model and a LCFT with $c=0$, which among other 
things implies that the dynamic exponent $z=1$. Several FSS exponents of
expectation values have been computed in the stationary state but 
their identification with the scaling dimensions of the LCFT has not
been completed. The FSS exponents of the avalanche PDF have also been determined. 
Much is still to be understood in this model, such as correlation 
functions and its off-critical behavior. We hope to come back to these 
topics in a future publication.

\section*{Acknowledgements} This research is supported by the 
Australian Research Council, and by the Foundation `Fundamenteel
Onderzoek der Materie' (FOM). JdG thanks the Australian National
University where parts of this work were done. VR has been financially
supported by ARC IREX and thanks the Einstein Centre of the Weizmann
Institute and SISSA for hospitality. We thank M. T. Batchelor,
E. Domany, M. Flohr, S. Franz, H. Hinrichsen, Y. Kafri, D. Mukamel,
G. Mussardo, A. Nersesyan, A. Owczarek, Yu. Stroganov, A. Vespignani
and O. Warnaar for discussions.

\end{document}